\documentclass[aps,prl,twocolumn]{revtex4}
\input epsf
\usepackage{amsbsy}
\usepackage{latexsym}
\usepackage{color}
\usepackage{graphicx}
\usepackage{psfrag}

\usepackage{dcolumn}
\usepackage{amssymb}
\usepackage{amsmath}
\usepackage{epsf,epsfig}
\usepackage{color}
\usepackage{graphics,psfrag}
\usepackage{graphicx,psfrag}

\begin{document}
\title{Nonlinear wavelength selection in surface faceting under electromigration
}
\author{Fatima Barakat, Kirsten Martens, and Olivier Pierre-Louis}
\affiliation{Laboratoire de Physique de la Mati\`ere Condens\'ee et Nanostructures,\\
Universit\'e  Lyon 1, 43 Bd du 11 novembre, 69622 Villeurbane, France.\\
}
\date{\today}

\begin{abstract}
We report on the control of the  faceting of crystal surfaces by means of surface electromigration. 
When electromigration reinforces the faceting instability, 
we find perpetual coarsening with a wavelength increasing as $t^{1/2}$.
For strongly stabilizing electromigration, the surface is stable.
For weakly stabilizing electromigration, a cellular pattern is obtained,
with a nonlinearly selected wavelength. The selection mechanism is not caused
by an instability of steady-states, as suggested by previous works in the literature.
Instead, the dynamics is found to exhibit coarsening {\it before} reaching a continuous family of stable non-equilibrium steady-states.
\end{abstract}

\maketitle

In non-equilibrium conditions,
crystal surfaces undergo various instabilities leading to micro or nanostructures.
The ultimate faith of these topographical structures
is governed by the nonlinear dynamics of the surface.
Understanding the nonlinear processes at play is therefore 
important for many applications where 
one wishes to design structures with a given wavelength.
However, the complexity of nonlinear wavelength
selection mechanisms have hindered their control,
and strategies to achieve predictive design are still poorly understood.
A possible strategy is
to control existing instabilities by means of an external field. 
Here we explore the control of the faceting instability with
electromigration. 
The faceting instability is the decomposition of a surface
with a given average orientation into neighboring facetted orientations~\cite{Chame1996,Kim2009,Jeong1995,DiCarlo1992,Liu1993}.
This instability is of thermodynamic origin, and is driven
by the reduction of the surface free energy.
The presence of an electric current
leads to a surface electromigration mass flux $J_E$. This mass flux
can stabilize or destabilize crystalline surfaces
depending on its orientation-dependence~\cite{Dobbs1994,Chang2006,Misbah2010}.
Many studies have been devoted to these two instabilities.
The combination of electromigration and anisotropy is already known
to lead to nontrivial dynamics of voids,
with the appearance of nontrivial orientations and oscillatory dynamics~\cite{Kuhn2005}.
In this letter, we show that a faceting instability can be
controlled by electromigration.
In the presence of destabilizing electromigration,
the faceting instability is found to be reinforced, and perpetual coarsening
with a wavelength increasing as $t^{1/2}$ is found.
Strongly stabilizing electromigration supersedes the faceting instability and 
stabilizes the surface. This result is similar 
to the stabilization of the elastic stress-induced Grinfeld instability by electromigration
discussed in Ref.~\cite{Tomar2010}.
Our most striking result appears in the presence of weakly stabilizing electromigration:
we provide converging analytical and numerical 
evidences of the existence of a continuous branch of stable periodic steady-states.
Since the dynamics exhibits a Lyapunov functional 
({\it i.e.}~an effective non-equilibrium energy functional 
which is decreased monotonically during the dynamics), these steady-states
can be characterized as {\it non-equilibrium meta-stable states}.
Furthermore, when starting from random initial conditions, 
a wavelength larger than the wavelength emerging 
from the linear instability is selected. 
However, the nonlinear wavelength selection mechanism is different 
from the known scenarios of interrupted coarsening discussed in Refs.~\cite{Politi1996,Danker2003,Politi2004},
or of secondary instabilities such as the Eckhaus instability~\cite{Eckaus1965}, 
which all emerge from instabilities of steady-states.

\begin{figure}
\centering
\includegraphics[width=0.8\columnwidth,clip]{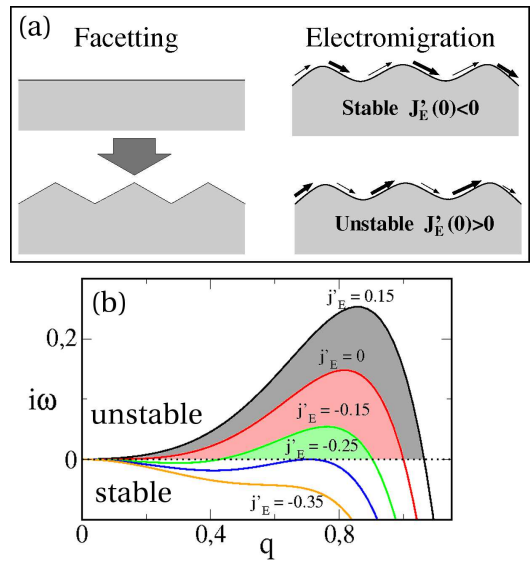}
\caption{{\it Mechanism of the instability and dispersion relation:} (a) Schematics of the 
faceting and electromigration-induced instabilities. (b) Growth rate $i\omega$ of 
Fourier modes from linear stability analysis, see Eq.~(\ref{eq-lin-stab}).}
  \label{fig-lin-stab}
\end{figure}

\paragraph{Model --}
Combining previous works on faceting~\cite{DiCarlo1992,Liu1993} and on electromigration-induced
instabilities~\cite{Dobbs1994,Chang2006,Misbah2010}, we write down a phenomenological nonlinear
equation which governs surface dynamics. 
We use a one-dimensional model for the crystal surface height profile $h(x,t)$
and slope $\phi(x,t)=\partial_xh(x,t)$.
The average orientation $\phi=0$ of the surface is assumed to be unstable, and to decompose
into facets of slopes $\phi=\pm 1$. For the sake of simplicity,
we consider a Ginzburg-Landau-like orientation dependent energy:
\begin{equation}
 \mathcal{F}[\phi]= \gamma \int dx \left[-\frac{\phi^2}{2}+\frac{\phi^4}{4}+\frac{\epsilon^2}{2}(\partial_x\phi)^2\right],
\label{e:free_energy}
\end{equation}
where $\gamma$ is a typical surface energy scale. 
The last term on the r.h.s. accounts for a curvature energy cost, 
which regularizes facets~\cite{DiCarlo1992,Liu1993} at the small lengthscale $\epsilon$, 
and allows one to write down local dynamics. 
From Eq.(\ref{e:free_energy}), the local chemical potential is derived as
\begin{align}
\mu=\Omega \frac{\delta\mathcal{F}[\phi]}{\delta h}
=\gamma\Omega\partial_x\left[\epsilon^2\partial_{xx}\phi+\phi-\phi^3\right].
\end{align}  
where $\Omega$ is the atomic area.
We consider surface-diffusion limited dynamics, so that chemical
potential gradients induce a surface mass flux $J_\mu=-D_L\partial_x\mu/k_BT$, 
where $k_BT$ is the thermal energy.
In addition, we assume that an external electric current is applied to the crystal,
leading to an orientation-dependent electromigration surface mass flux $J_E(\phi)$.
Expanding the electromigration current for small slopes, we write $J_E(\phi)\approx J_E(0)+\phi J'_E(0)$.
The slope-dependence of $J_E$ may either be caused by the anisotropy of
surface diffusion~\cite{Stoyanov1990,Dobbs1994}, or of the migration force~\cite{PierreLouis2006}.
As shown in Fig.\ref{fig-lin-stab}(a), electromigration is destabilizing when $J'_E(0)>0$, and
stabilizing when $J'_E(0)<0$. This instability is the subject of
a large literature, and has been the basis of the interpretation
of the instabilities observed on semiconductors~\cite{Latyshev1989}
and on metals~\cite{Johnson1938}.
Mass conservation reads $\partial_th=-\Omega\partial_x[J_\mu+J_E]$, and leads to an evolution
equation for the surface height:
\begin{eqnarray}
\label{heigt-dyn}
\frac{ \partial_t h }{\Omega}&=& D_L\frac{\Omega\gamma}{k_BT}\partial_{xx}\left[\epsilon^2\partial_{xxxx}h
+\partial_{xx}h-\partial_x(\partial_x h)^3\right]
\nonumber \\  
&-&J_E'(0)\partial_{xx}h\;,
\end{eqnarray}
where we have assumed for simplicity that $D_L$ does not depend on the orientation.
Normalizing $x$ with $\epsilon$, $t$ with $k_BT\epsilon^4/\Omega^2\gamma D_L$, and defining
the dimensionless parameter $j'_E=J'_E(0)k_BT\epsilon^2/\Omega\gamma D_L$, we may write a one-parameter
equation for the slope:
\begin{eqnarray}
\label{slope-dyn}
 \partial_t \phi =\partial_{xxxx}\left[\partial_{xx}\phi+\phi-\phi^3\right]-j'_E\partial_{xx}\phi\;.
\end{eqnarray}
The relaxation part of Eq.(\ref{slope-dyn}), first derived in Ref.~\cite{Liu1993},
is expected to give rise to perpetual logarithmic coarsening as discussed in Ref.~\cite{Hausser2009}.
The electromigration part has been derived and used in many papers, see Ref.~\cite{Misbah2010} for a review.
When electromigration is destabilizing  ($j'_E>0$), powerlaw coarsening
is expected. Here, we claim that the combination of the two physical processes, 
faceting and electromigration, gives rise to novel dynamics.

\paragraph{Linear stability analysis --}
Let us start the study of Eq.(\ref{slope-dyn}) by means of a linear
stability analysis. The growth rate $i\omega$ of a Fourier mode 
$\sim {\rm e}^{i\omega t+qx}$ of wavelength $\lambda=2\pi/q$ 
reads:
\begin{equation}
 i\omega = -q^6+q^4+j_E' q^2\;.
\label{eq-lin-stab}
\end{equation}
For $j_E'=0$, all wavelengths larger than $2\pi$ are unstable.
For $j'_E>0$ the range of unstable wavelengths is still infinite, but extends to smaller
and smaller wavelengths: $\lambda>\lambda_+=2\pi/[1/2+(1/4+j'_E)^{1/2}]^{1/2}$.
When electromigration is weakly stabilizing  $0>j'_E>-1/4$, there is a finite
range of unstable wavelengths $\lambda_-<\lambda<\lambda_+$ 
where $\lambda_-=2\pi/[1/2-(1/4+j'_E)^{1/2}]^{1/2}$.
Finally, for strongly stabilizing electromigration, when $j'_E<-1/4$, 
the surface is linearly stable. When the surface is unstable, the wavelength
$\lambda_m=2\pi 3^{1/2}/[1+(1+3j'_E)^{1/2}]^{1/2}$ of the fastest
growing mode is expected to emerge from random initial conditions
(where all modes are present).

\paragraph{Lyapunov functional --}
The linear stability analysis predicts the existence
or not of an instability from small initial perturbations.
However, the amplitude of the unstable modes increases
exponentially fast, and the unstable front enters into the nonlinear
regime where the $\phi^3$ term in Eq.(\ref{slope-dyn})
is not negligible anymore. While the analytical study of the nonlinear regime is in general
delicate, the existence of an effective energy functional $\mathcal L$, sometimes
called a  Lyapunov functional, usually greatly simplifies the analysis.
Such a functional is defined as a quantity which 
decreases monotonously during the dynamics. 
The evolution equation (\ref{slope-dyn}) actually exhibits a Lyapunov
functional (in normalized coordinates):
\begin{eqnarray}
\label{eq-lyapunov}
 \mathcal{L}[h]= \mathcal{F}[\partial_x h]-\frac{j_E'}{2}\int dx\; h(x)^2.
\end{eqnarray}
Indeed, one may write
$\partial_t h = \partial_{xx} [\delta \mathcal{L}/{\delta h}]$,
leading to:
\begin{align}
\partial_t\mathcal{L}= -\int dx\left[\partial_x(\delta \mathcal{L}/\delta h)\right]^2\leq 0.
\label{e:variational_dynamics}
\end{align}

\paragraph{Steady-state branches --}
The space of all possible configurations for the surface profile
is very large, and it is impossible to calculate the value of 
$\mathcal{L}$ for all configurations. 
As a consequence, we would like to reduce this space to a relevant subset
of shapes which is simpler to explore. 
A powerful approach along this line is to study the steady-states $\phi_0(x)$,
which are the solutions of $\partial_t\phi_0=0$.
Using Eq.(\ref{e:variational_dynamics}), one has $\partial_x(\delta \mathcal{L}/\delta h)=0$,
leading to
\begin{align}
\partial_{xx}\left[\partial_{xx}\phi_0+\phi_0-\phi_0^3\right]-j'_E\phi_0=0.
\label{e:steady_states}
\end{align}
The central idea motivating the study of steady-states 
is the existence of a separation of timescales,
where the shape first relaxes rapidly towards periodic steady-states.
Then, these periodic solutions may exhibit an instability occurring at
longer timescales, leading for example to a coarsening
process~\cite{Politi2004}, or to the Eckhaus instability~\cite{Eckaus1965}. 

\begin{figure}
\centering
\includegraphics[width=\columnwidth,clip]{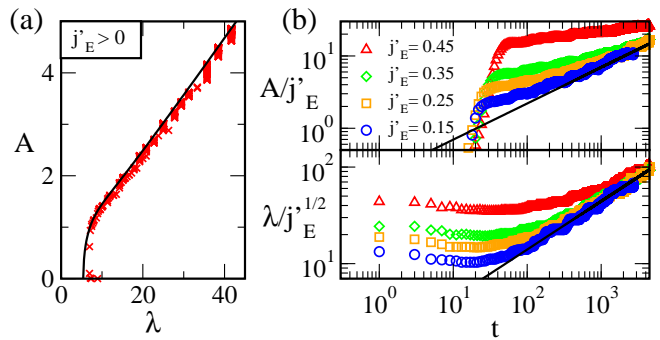}
\caption{{\it Perpetual power-law coarsening for $j'_E>0$:} 
(a) The black line indicates the steady state amplitude $A$ versus wavelength $\lambda$ of
periodic steady-states for $j'_E = 0.45$. 
Crosses ({\color{red}$\times$}): full simulations from small random initial conditions. 
(b)  Rescaled Amplitude $A/j'_E$ and wavelength $\lambda/(j'_E)^{1/2}$ 
as function of time for starting from small random initial conditions.
The solid lines are the powerlaws discussed in the text.}
  \label{fig-large-je}
\end{figure}

The steady-states profiles are obtained numerically, starting with 
a small amplitude sinusoidal perturbation $\sim \sin(2\pi x/\lambda)$ 
with one period in a box of width $\lambda$\footnote{We use an finite difference explicit
scheme.}. Following the linear stability analysis,
the perturbation may grow or decay. When it grows, it reaches a finite amplitude
steady-state where it stops. As shown in Fig.\ref{fig-large-je} for $j'_E>0$,
the steady-state  exhibits a monotonously increasing amplitude $A$ as a function
of the wavelength $\lambda$. 
Asymptotically for large $A$ and $\lambda$,
we expect the dominant term among the linear
terms to be the one with the fewest derivatives, i.e. $j'_E\phi_0$,
and this term has to balance the nonlinear term $\partial_{xx}\phi_0^3$,
leading to $A\sim (j'_E)^{1/2}\lambda$. The numerical determination
of the steady-state branch actually indicates that:
$A\approx 0.16 (j'_E)^{1/2} \lambda$.

\begin{figure}
\centering
\psfrag{L1}[b][b][1.2]{$\mathcal{L}/\lambda$}
\psfrag{L2}[b][b][0.8]{$\mathcal{L}$}
{	
\includegraphics[width=\columnwidth,clip]{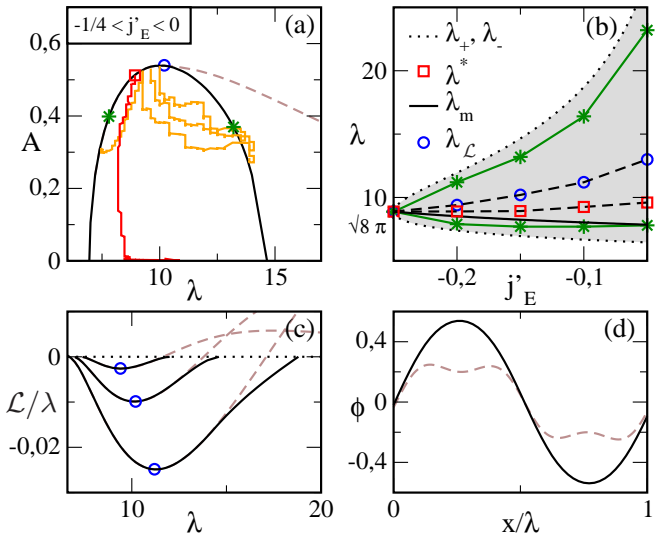}
}
\caption{{\it Nonlinear wavelength selection for $-1/4<j'_E<0$}: 
(a) The (black) solid and (brown) dashed lines indicate the steady state amplitude $A$ 
versus wavelength $\lambda$ for $j'_E = -0.15$. 
 Red line: full dynamics from small random 
initial conditions in an extended system. 
Orange lines: dynamics starting from slightly perturbed periodic steady-state
in an extended system. Green stars indicate theoretical limit of
stability of the steady-state branch.
(b) 
The shaded region corresponds to the linearly unstable region.
Green stars: limits of steady-state stability. 
(c) Lyapunov functional density $\mathcal{L}/\lambda$ evaluated from the steady state profiles. 
(d) Steady state profiles numerically obtained for the two branches at $j'_E=-0.15$.}
\label{fig-small-je}
\end{figure}

When $-1/4<j'_E<0$, we obtain a bell shape branch 
connecting the two marginal points $(\lambda=\lambda_-,A=0)$, and $(\lambda=\lambda_+,A=0)$,
as shown in Fig.\ref{fig-small-je}(a). 
In addition, using suitable initial conditions,
we find another steady-state branch, which
emerges from the primary bell-shape branch. The typical
steady-state profiles are shown on Fig.\ref{fig-small-je}(d).
Other branches could exist, and we have not tried to provide a complete
analysis of all possible steady-state branches.
However, and as discussed in the following, the study of the main
bell-shape branch seems to be sufficient to account for the main features of the 
full dynamics starting from small random perturbations.

The numerical calculation of ${\cal L}$ along the steady-state branches 
shows that ${\cal L}$ decreases monotonously with $\lambda$ for $j'_E>0$,
while it exhibits a minimum for $\lambda=\lambda_{\cal L}$ when $-1/4<j'_E<0$.
At this point, it is tempting to speculate that the dynamics will simply
follow the gradient of ${\cal L}$ along the steady-state branches
\footnote{More precisely, since $\lambda$ varies, we calculate the density ${\cal L}/\lambda$.},
leading to infinite coarsening when $j'_E>0$ and interrupted coarsening
at $\lambda=\lambda_{\cal L}$ for $-1/4<j'_E<0$. We shall see in the 
following that some of these speculations are actually wrong.

\paragraph{Steady-state stability --}
In order to analyze the dynamics more carefully, 
we shall investigate the stability of steady-states
with respect to small perturbations  $\phi_1(x)=\phi(x)-\phi_0(x)\ll \phi_0(x)$.
The slope variation $\phi_1$, leads to the following 
variation of ${\cal L}$:
\begin{align}
{\cal L}_1=
\frac{1}{2}\int dx \left[\phi_1^2(3\phi_0^2-1)+(\partial_x\phi_1)^2-j'_Eh_1^2\right].
\label{e:variation_L}
\end{align}
where $\partial_x h_1=\phi_1$.
Physically relevant steady-states must be stable with respect to perturbations
with wavelengths smaller than their periodicity, and in general, we expect that
the most dangerous modes 
are long-wavelength perturbations, as pointed out e.g. in Ref.~\cite{Politi2004}.
In the long wavelength limit, where the perturbation wavelength is much
larger than the periodicity of the steady-state $\phi_0$, one may simply replace
$\phi_0^2(x)$ by its average over one period $\langle \phi_0^2\rangle$.
Thus, since it has constant coefficients,  Eq.(\ref{e:variation_L})
is now diagonal in Fourier space, and
this suggests a simple stability criterion:
\begin{align}
(3\langle\phi_0^2\rangle-1)q_1^2+q_1^4-j'_E>0.
\label{e:stability_long_wavelength}
\end{align}
As a consequence, long wavelength modes $q_1\rightarrow 0$ are always unstable
in the case of destabilizing electromigration $j'_E>0$.
This is in agreement with the above speculation of perpetual coarsening.
In addition, we may also gain information
about the coarsening exponent. Indeed, at long wavelengths
we have ${\cal L}_1\approx -\int dx \,j'_E h_1^2/2$, and as a consequence
one has $\partial_th_1\approx -j'_E\partial_{xx}h_1$.
This relation provides a link between lengthscales $x$, assumed to be $\sim \lambda$,
and timescales $t$, leading to $\lambda \sim (j'_E) ^{1/2}t^{1/2}$. 
Using the previously derived linear relation between $\lambda$ and $A$, we also find
$A\sim j'_Et^{1/2}$.

When $-1/4<j'_E<0$, from Eq.(\ref{e:stability_long_wavelength}),
the steady-states should be stable for all $q_1$ when
\begin{align}
\langle\phi_0^2\rangle>\frac{1}{3}[1-2(-j'_E)^{1/2}].
\label{e:stabil_WDE}
\end{align}
This criterion indicates that the upper part of the branch between the green stars
in Fig.\ref{fig-small-je}(a,b) should be stable, while the lower parts
close to $\lambda_\pm$ should be unstable
\footnote{Due to numerical difficulties in the calculation of the
steady-state branch for $\lambda$ close to $\lambda_+$ when $0>j'_E\geq -0.1$,
the upper stability limit in Fig.3(b) was obtained from an interpolation for $j'_E=-0.1,-0.05$.}.
Such a result suggests that the system is stuck once the dynamics 
hits the stable part of the steady-state branch, and as a consequence,
it cannot evolve towards the minimum of ${\cal L}$ at $\lambda=\lambda_{\cal L}$.

\begin{figure}[h]
\centering
\includegraphics[width=\columnwidth,clip]{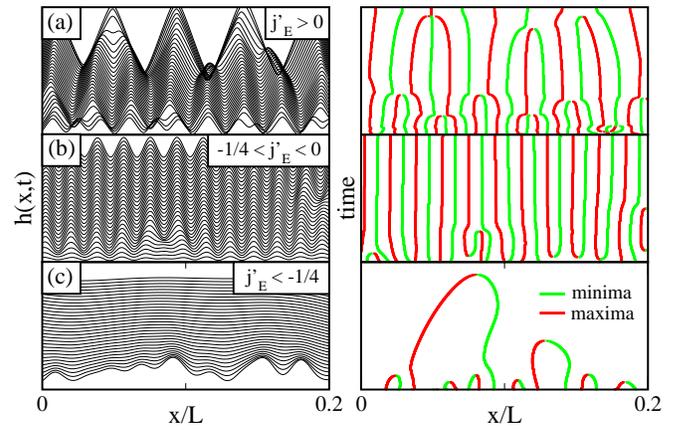}
\caption{{\it Full dynamics and coarsening:} 
Left: surface height $h(x,t)$ as a function of space for different 
times in the three different regimes: (a) $j_E'=0.45$, (b) $j_E'=-0.15$, 
(c) $j_E'=-0.3$. Right: Corresponding spatiotemporal portrait of the extrema of the height profile.}
  \label{fig-dynamics}
\end{figure}

\paragraph{Full dynamics --}
These results are confirmed by the full numerical
solution of Eq.(\ref{slope-dyn}) starting from small random 
perturbations of a flat state in a system of size $L=500$.
The results are shown in Fig.\ref{fig-dynamics}.
First, and as expected, we find stable dynamics for 
$j'_E<-1/4$. 

For destabilizing electromigration $j'_E>0$, perpetual coarsening
is found, as shown on Fig.\ref{fig-dynamics}(a), and after a transient
related to the linear instability, the dynamics
follows the steady-state branch in the $(\lambda,A)$ plane,
as shown in Fig.\ref{fig-large-je}(a). We also confirm 
the scaling laws given above in Fig.\ref{fig-large-je}(b), and find the prefactors:
$\lambda\approx 1.4 (j'_E)^{1/2}t^{1/2}$, and $A\approx 0.22 \,j'_E\,t^{1/2}$.
The ratio of these prefactors is in perfect agreement
with the asymptotic behavior of the steady-states discussed above $A\approx 0.16(j'_E)^{1/2}\lambda$.

For weakly stabilizing electromigration $-1/4<j'_E<0$,
the dynamics starts with a rapid increase of the amplitude
due to the linear instability. Then, the average wavelength
increases  {\it before} the dynamics hits
the steady-state branch at $\lambda=\lambda^*$, as shown in Fig\ref{fig-small-je}(a). 
When the dynamics reaches the steady-state branch,
it stops, as expected from the above prediction of steady-state
stability. As a consequence, the system never reaches the 
minimum of ${\cal L}$ along the steady-state branch.

In order to check further the stability of the steady-state
branch, we have performed simulations 
in boxes of width around $80$ periods starting
with periodic steady-states with small perturbations. These simulations confirm the stability
of the upper part of the branch, and the instability of the lower part, in quantitative
agreement with the condition (\ref{e:stabil_WDE}).
As seen in Fig.\ref{fig-small-je}(a), the instability does not lead to a trajectory
of the system along the steady-state branch in the $(\lambda,A)$ plane. 
Instead, the trajectory escapes from the branch
and returns to it, stopping in the same region
as the dynamics from flat initial conditions. Fig.\ref{fig-small-je}(b)
summarizes the evolution of the different lengthscales as a function
of $j'_E$.

\paragraph{Conclusions and perspectives --}
In summary, we have studied the control of the faceting instability
by means of electromigration.
For strong stabilizing electromigration, the surface is stable.
Under weakly stabilizing electromigration, the surface
exhibits a periodic cellular structure with a nonlinearly selected wavelength.
When electromigration is destabilizing, 
perpetual coarsening is found with a coarsening exponent $1/2$.

In the case of weakly stabilizing electromigration, we find a continuous family of stable
periodic steady-states. These states can be interpreted as
 non-equilibrium metastable states, where metastability
is here defined as local stability with respect to a non-equilibrium 
Lyapunov functional playing the role of an effective energy.

In the literature, several cases of nonlinear wavelength selection on
crystal surfaces have been interpreted as interrupted coarsening, 
such as for mound growth~\cite{Politi1996}, atomic step meandering~\cite{Danker2003},
and ion sputtering~\cite{MunozGarcia2010}.
However, the nonlinear wavelength selection scenario presented in this Letter
does not correspond to interrupted coarsening, as defined e.g.~in Ref.~\cite{Politi2004},
or to another instability of steady-states, such as in the Eckhaus instability~\cite{Eckaus1965}. 
Indeed,  we only observe a small amount of coarsening
{\it before} the dynamics hits the steady-state branch.
Once the system has reached a steady-state, it is stable
and the evolution stops.

These results may provide hints to understand other nonlinear wavelength
selection scenarios obtained in the literature, such as in the combination
of growth and faceting as discussed in Ref.~\cite{Golovin2001,Savina2003} (where nonlinear wavelength selection
occurs between a coarsening regime, and a chaotic regime).
More generally, we hope that our work 
will provide milestones towards the novel methods to control the size of nano-structures emerging
from surface morphological instabilities. 

\acknowledgments
KM was supported by the Marie Curie FP7-PEOPLE-2009-IEF program.

\bibliography{refs}

\end{document}